# Measurement of single-cell elasticity by nanodiamond-sensing of non-local deformation

Yue Cui[1+], Weng-Hang Leong[1+], Chu-Feng Liu[1], Kangwei Xia[1], Xi Feng[1], Csilla Gergely[4], Ren-Bao Liu[1,2,3*], Quan Li[1,2,3*]

1. Department of Physics, The Chinese University of Hong Kong, Shatin, New Territories, Hong Kong, China

2. Centre for Quantum Coherence, The Chinese University of Hong Kong, Shatin, New Territories, Hong Kong, China

3. The Hong Kong Institute of Quantum Information Science and Technology, The Chinese University of Hong Kong, Shatin, New Territories, Hong Kong, China

4. Laboratoire Charles Coulomb, University of Montpellier, Montpellier, France

[+]These authors contributed equally: Yue Cui, Weng-Hang Leong

[*]Correspondence and requests for materials should be addressed to R.-B.L. (email: rbliu@cuhk.edu.hk) or to Q.L. (email: liquan@ cuhk.edu.hk)

# Abstract

Nano-indentation based on, e.g., atomic force microscopy (AFM), can measure single cell elasticity with high spatial resolution and sensitivity, but relating the data to cell mechanical properties depends on modeling that requires knowledge about the local contact between the indentation tip and the material, which is unclear in most cases. Here we use the orientation sensing by nitrogen-vacancy centers in nanodiamonds to chart the non-local deformation of fixed HeLa cells induced by AFM indentation, providing data for studying cell mechanics without requiring detailed knowledge about the local contact. The competition between the elasticity and capillarity on the cells is observed. We show that the apparent elastic moduli of the cells would have been overestimated if the capillarity is not considered (as in most previous studies using local depth-loading data). We also find reduction of both elastic moduli and surface tensions due to depolymerization of the actin cytoskeleton structure. This work demonstrates that, under shallow indentation, the nanodiamond sensing of non-local deformation with nanometer precision is particularly suitable for studying mechanics of cells.

# Introduction

Cells exhibit characteristic mechanical properties in different physiological and pathological processes, such as cell division [1], motility [2,3], differentiation [4,5], cancer development [6] and metastasis [7,8]. Atomic force microscopy (AFM) indentation is widely adopted to measure cell elasticity, featuring nanoscale resolution and piconewton sensitivity. However, the AFM-based approach relies on local depth-loading measurement at the contact region and therefore on detailed indentation parameters (e.g., tip geometry [9], indentation speed [10], contact point [11]). The requirement of precise knowledge about local contact [12] and the complexity of cell mechanical properties (e.g., viscosity [13], adhesion [14], surface morphologies [15]) make the interpretation of indentation data complex and often ambiguous.

Measuring the non-local response of cells (surface profiles) to local AFM indentation is a possible solution to this outstanding problem. The non-local response is independent of the local indentation parameters and therefore can be employed to reduce the ambiguity in relating the deformation data to the mechanical properties. In fact, optical imaging of non-local deformation profiles has been applied to soft polymers, leading the discovery of the interplay between capillary forces and elasticity [16–19]. This method of non-local deformation measurement, however, does not have sufficient resolution and precision for studying cells. The elastocapillary effects, i.e., the competition between elasticity and capillarity, become relevant in cells usually at sub-micron length scales [15,17], which is often unresolvable in deformation induced by microspheres in optical imaging. Moreover, sub-micron shallow indentation is desirable for cellular studies because it is less invasive [6], but the required nanometer precision of the deformation measurement is beyond the capability of optical imaging methods.

The nanodiamond deformation mapping we developed earlier [20] provides a solution. Electron spins of nitrogen-vacancy (NV) centers in diamond are promising for biomechanical sensing [21–26]. The vector magnetometry using the dependence of NV center spin resonances on the magnetic field along the NV axis can be used to measure the orientation of nanodiamonds (NDs) docked on a surface and hence the surface deformation [20]. With such a technique, the effect of surface tension in polymer has been recently disclosed by reconstructing the non-local deformation under AFM indentations. The ND-based deformation mapping features a resolution ~200 nm and a precision ~5nm [20], which meets the requirements for studying mechanics of cells.

Here we apply the ND deformation mapping approach to study the intrinsic mechanical properties of cells and their changes under controlled structural manipulations. For a supervised methodological demonstration in this first-stage proof-of-concept exploration, we choose fixed cells to avoid possible complications due to active cellular changes such as internal motion and/or cell cortex modifications, etc. [27]. We use AFM tips to induce shallow indentation of fixed HeLa cells. The deformation away from the tips is measured by ND orientation sensing and the data are fitted by a linear elastic model with the surface tension effect included. We show that the elastic modulus and the surface tension can be simultaneously evaluated using the deformation data, which enables a more precise evaluation of the cell stiffness. The stiffness would be significantly overestimated if the apparent elastic modulus is used, as in conventional methods that employ classical contact models without considering the surface tension [28]. Then we measure the mechanical properties of cells treated with an actin-interfering drug (Cytochalasin D) and compare the data with those from control samples. Both the elasticity and the surface tension of cells are reduced by the treatment, which we ascribe to the structure of actin cortex being disrupted by Cytochalasin D. This effect of drug treatment on cell elasticity would be largely underestimated if the capillarity is ignored in modelling the deformation data. These results constitute the first unambiguous experimental observation of the elastocapillary effect of AFM indentation on cells.

## Results

**Scheme and numerical simulation**

The local depth-loading profile in the indentation measurement depends on the features of the contact between the point and sample, the uncertainty in which would lead to ambiguity in data interpretation. In contrast, the deformation at locations away from the indentation point (i.e., non-local deformation), according to the Saint-Venant's principle in linear elastic theory [29], is independent of the contact details but rely solely on the intrinsic mechanical properties of the materials. Therefore, monitoring the non-local deformation can reveal the intrinsic mechanical properties of soft, biological materials, irrespective of local uncertainties.

In an indentation measurement carried out in liquids (Fig. 1a), at the material-liquid interface there is competition between two intrinsic mechanical properties of the material, i.e., the capillary force (characterized by surface tension $\tau_0$) and the elastic response (characterized by elastic modulus $E^*$, which includes contributions from both the shear and bulk moduli, as usually used in contact mechanics called as effective Young's modulus [12]). This so-called elastocapillary effect [17] is characterized by the elastocapillary length $s \equiv 2\tau_0/E^*$ [30]. The elastocapillary length $s$ is the size of a deformation area in which the surface tension force and the bulk elastic force are comparable (Fig. 1a). In the deformation regime close to the indentation point with $\rho \ll s$ ($\rho$ is the distance from the indentation point), the capillary force dominates, while for $\rho \gg s$ the bulk elasticity is the governing contributor.

For a typical soft, biological material, such as the cell surface/sub-surface (composed of lipid membrane and underlying actin cortex with a typical thickness of 100 to 1000 nm [12]), the elastocapillary length $s$ is in the order of 1 μm [31]. For simultaneously evaluating the surface tension and the bulk elasticity, it is important to measure them at sub-micron scale where the competition is most relevant. On the other hand, shallow indentation is desired for a less invasive measurement, and also to minimize the contributions from internal cytosol or other structures of the cell [12]. The shallow indentation (with depth of 200 to 700 nm) is expected to result in a deformation in the range of 40 to 170 nm at ~ 1 μm away from the indentation point (see Supplementary

Fig. 2 for the estimation). Consequently, a method that can map the non-local deformation at $\rho \sim 1$ μm with submicron lateral resolution and nanometer precision in the vertical deformation is required to simultaneously evaluate the capillarity and elasticity of cells.

In this regard, ND based orientation sensing integrated with AFM indentation is an ideal solution. Therefore, we docked NDs on the cell surface under the AFM indentation (Fig. 1b) and we monitored the ND orientation based on vector magnetometry [20,22]. The ND orientation change upon AFM indentation is described by its rotation axis $\mathbf{u}(\theta, \phi)$ and rotation angle $\chi$ (Fig. 1c), which is used to reconstruct the non-local deformation through the geometric relation $\chi \approx \partial_\rho z$ for an isotropic deformation, where $z(\mathbf{\rho})$ is the vertical deformation at the location with displacement $\mathbf{\rho}$ from the indentation point (details in Methods).

We mapped the non-local deformation of cell upon shallow indentation (200-700 nm) by AFM (~50 nm tip radius) at $\rho > 300$ nm. The non-local deformation profile $z(\rho)$ is well approximated as the deformation upon point loading, which confirms that the non-local data is free of uncertainties in the contact model. Such $z(\rho)$ profile is a universal Hajji curve with only two free parameters, the typical length $s$ and the overall amplitude $z_P \equiv P/\tau_0$ proportional to the indentation force $P$ (see Eq. (2) in Methods) [30]. Consequently, the rotation angle $\chi$ of a docked ND, when normalized by $z_P/s$, is a universal function of $\rho/s$ (see Eq. (3) in Methods), as shown in Fig. 1d. Under the asymptotic conditions, the rotation angle is approximated to either $\chi \propto \rho^{-2}$ for $\rho \gg s$ (blue dashed line in Fig. 1d), recovering to the Hertz-Sneddon model [28] in the elasticity-dominated situation; or $\chi \propto \rho^{-1}$ for $\rho \ll s$ (green dashed line in Fig. 1d), obeying the Young-Laplace equation in the capillary-dominated situation. Fitting the curve $\chi(\rho)$ in the transition region ($\rho \sim s$), one can get the deformation amplitude $z_p \equiv P/\tau_0$ and the typical length $s = 2\tau_0/E^*$, and consequently, with the loading force $P$ known, get both the elastic modulus $E^*$ (the elasticity) and the surface tension $\tau_0$ (the capillarity).

**Non-local deformation reconstruction of fixed HeLa cells**

As a first demonstration of the method in cellular study, we chose fixed cells to avoid complications in data interpretation (e.g., cell structural changes from intracellular activities such as cell cortex evolution). Figures 2a and b respectively show the AFM topography and the confocal fluorescence images of a fixed HeLa cell with NDs anchored on the cell membrane (for cell culture and sample preparation, see Methods). The white dashed lines mark the boundary of the cell, and the bright spots in Fig. 2b indicate the fluorescence signals of NDs. Homogeneity of the mechanical response of the cell cortex is suggested by the similar depth-loading profiles acquired at different locations on the fixed cell sample (see Supplementary Fig. 3), which is consistent with literature [32]. By locating a single ND probe, we performed sequential AFM indentations at 21 different locations on the cell membrane in the proximity of the ND (highlighted by the green hexagon and the white arrow in Figs. 2a and b), respectively, under a constant loading force ($P = 13$ nN, see Methods for the AFM indentation). For an approximately homogeneous cell cortex in a fixed cell, the rotation of a single ND due to indentation of the AFM tip at different locations is equivalent to the rotation of multiple NDs at different locations due to indentation as a fixed position. The optically detected magnetic resonance (ODMR) spectra of the NDs under two different magnetic fields were collected with and without AFM indentation. One set of such ODMR spectra is shown in Fig. 2d (all others in Supplementary Fig. 4). The rotation of the ND was extracted from the resonance frequency shift with and without AFM indentation (Fig. 2d). The cell sample remained intact after the indentation, as suggested by the almost identical ODMR spectra (see Supplementary Fig. 4a & b) taken after the indentation experiments (AFM tip withdrawn from the sample) at different location of the cells. The rotation of the ND at $\boldsymbol{\rho} = (\rho, \phi_p)$ is characterized by the rotation axes and the rotation angle $\chi$ (Fig. 2c). Figure 2e shows the three-dimensional reconstruction of the cell membrane deformation using the gradient obtained from the ND rotation data (see Methods). The membrane deformed axisymmetrically due to the AFM indentation by an axisymmetric tip (see the scanning electron microscope (SEM) image in Supplementary Fig. 1). As shown in Figs. 2f & 2g, the rotation axes $\mathbf{u}$ of the ND on the cell membrane approximately lied on the $x - y$ plane and the ND rotated toward the indentation location (i.e., $\mathbf{u} \perp \boldsymbol{\rho}$). The rotation angle $\chi$ of the ND as a function of distance $\rho$ is shown in Fig. 2h, which only depends on the magnitude of the displacement $\boldsymbol{\rho}$ and is insensitive to its direction within error. These results are consistent with the axisymmetric indention on the planarly homogenous surface.

**Cell mechanical properties measured by non-local deformation**

Figure 2h plots the rotation angle of the NDs on the cells as functions of $\rho$ (green triangles), which are well fitted by the Hajji function under the given constant force $P$ (green line). The elastic modulus and the surface tension of the cell are estimated as $E^* = 52 \pm 20$ kPa, and $\tau_0 = 22 \pm 7$ mN m$^{-1}$, respectively. The deduced deformation (green triangles) and the corresponding fitting result (green line) are shown in Fig. 2i. The experimental data deviate significantly from the simulation (grey dashed lines in Figs. 2h & i) with the conventional Hertz-Sneddon model that does not include the surface tension (measured indentation depth of $345 \pm 22$ nm (Supplementary Fig. 3c) is used, details in Methods). This deviation shows the importance of the elastocapillary effect on shallow AFM indentation on cells.

Considering possible differences between individual cells, the non-local deformation measurements were performed on more cells (for details of the measurements see Supplementary Fig. 5 and 6). The measured rotation axes **u** of the NDs on all cells are shown in Fig. 3a, all with **u** ⊥ **ρ** within the measurement errors ($\sigma_\theta = 14°$ and $\sigma_{\Delta\phi} = 41°$, for the polar and azimuthal angles, respectively), which is consistent with the axisymmetric indentation. Figure 3b shows the measured rotation angle $\chi$ (color triangles) normalized by the constant $z_P/s$ as functions of the rescaled distance $\rho/s$, where the characteristic lengths $z_P$ and $s$ are obtained by fitting $\chi$ using the Hajji function. The scaled experimental data for all cells fall into a single Hajji function (dashed lines). Under the given loading force (see Supplementary Figs. 3b and 6b), the elastic modulus and the surface tension of the cells are deduced (Fig. 3c) as $E^* = 97 \pm 67$ kPa (median, 77 kPa, $n = 21$) and $\tau_0 = 8 \pm 6$ mN m$^{-1}$ (median, 8 mN m$^{-1}$, $n = 21$). It is noted that the surface tension of the fixed Hela cells we measured ($\sim 8$ mN m$^{-1}$) is higher than the reported values of live HeLa cells ($\sim 1$ mN m$^{-1}$) [6,7]. The difference may be attributed to the increased protein crosslinking level in cells during the fixation by PFA [35–37].

As a comparison, we also calculated the apparent elastic modulus $E_a^*$ by fitting the local depth-loading curves with the conventional Hertz-Sneddon model, which does not include the surface tension (see Methods). The obtained apparent elastic modulus ($E_a^* = 308 \pm 129$ kPa, median, 298 kPa, $n = 21$) is consistent with the literature on fixed HeLa cells [38,39], but is significantly larger than the values deduced from the non-local measurement (Fig. 3c). Our results suggest that the surface tension contributes significantly to the mechanical responses upon AFM indentation and affects the evaluation of the elastic modulus of cells. Considering the surface tension effect [31], it would be interesting to revisit some previous discoveries on cell mechanics that were performed by local depth-loading measurement, such as the effects of sub-structure heterogeneity [32] proposed by the observation that the apparent elastic modulus decreases with increasing the indentation depth.

**Change of cell mechanics upon disruption of actin filament networks**

To gain further insight into how actin cortex contributes to cell mechanics, we investigated the effect of Cytochalasin D (CytD) treatment on cell elasticity and surface tension. CytD disrupts the actin filament networks and binds to the filament ends to inhibit the polymerization of actin monomers [40]. To visualize the effect of CytD on actin cortex, F-actin in cells were stained using phalloidin. Figure 4a shows the fluorescence image of a labeled HeLa cell, where the F-actins form a fibrous meshwork constituting the actin cortex. After a 40 min. treatment with 5 µg mL$^{-1}$ CytD (for details see Methods), the actin cortex organization was modified from meshwork to short actin oligomers (focal accumulations) as shown in Fig. 4b. We measured the mechanical properties of CytD-treated fixed HeLa cells using the ND orientation sensing method (see Supplementary Figs. 7 to 9). As illustrated in Figs. 4c and d, in comparison to results for untreated cells, the CytD treatment caused a slight decrease of the surface tension (8 mN m$^{-1}$ to $3 \pm 3$ mN m$^{-1}$, median, 2 mN m$^{-1}$, $n = 20$; Welch's $t$-test $P = 0.008$) and a significant decrease of the elastic modulus (97 kPa to $20 \pm 16$ kPa, median, 14 kPa; $P = 5.7 \times 10^{-5}$). The reduction of the elastic modulus by the CytD treatment indicates that the cytoskeletal network is the main structural contributor to the measured cell stiffness, as reported in previous works [41,42]. The slight decrease of the surface tension could result from the depolymerization of actin filaments, while the reduced number of filaments leads to smaller surface tension at the cell-liquid interface [35,37]. We further investigated how the CytD induced cell stiffness change would manifest itself in the non-local measurement (Fig. 4d) and in the conventional assessment by local AFM indentation without considering the surface tension effect (Fig. 4e). The elastic moduli of CytD treated cells, as obtained from the non-local deformation measurement, are found to be reduced to ~1/5 of those of the untreated cells (from 97 kPa to 20 kPa). In comparison, the apparent elastic moduli deduced from conventional local AFM indentation are found to be decreased to ~1/2 of those untreated (by CytD) (from 308 kPa to $123 \pm 71$ kPa, median, 107 kPa, $n = 20$; $P = 4 \times 10^{-6}$). The results suggest that the drug effect/structure change of cell mechanics is significantly underestimated if the surface tension is neglected.

**Conclusions**

By combining AFM indentation and ND orientation sensing, we demonstrated high-precision, non-local deformation measurement of individual adherent mammalian cells upon shallow and hence less invasive indentation. The non-local nature of the method enables the mechanical properties to be unambiguously interpreted without requiring detailed knowledge of the local indentation (e.g., tip geometry, indentation speed, contact point, etc.). A quantification method is developed to enable simultaneous evaluation of the elastic modulus and surface tension of cell surface/subsurface structures (membrane and cortex). The nanometer precision of the non-local measurement (see Supplementary Figs. 10) allows detection of small deformation (being comparable to the sub-micrometer elastocapillary length of cells) upon shallow indentation, revealing the nontrivial effect of surface tension in evaluating cell mechanics by AFM indentation. Neglecting the surface tension effect leads to an overestimation of the apparent elastic modulus and an underestimation of cell stiffness variations due to cortical structure alteration (e.g., by drug treatment). The ND-based deformation sensing thus adds to the existing AFM techniques for cell mechanics study and for diagnostic/prognostic applications.

## Methods

**Experimental setup**

The measurements were carried out on a home-built confocal-AFM correlation microscope (for details see Supplementary Information of Ref. [20]). We used the laser scanning confocal system for imaging and ODMR measurements. The spins of NV centers were pumped by a 532 nm laser and manipulated by microwave delivered via a copper wire antenna (20 μm diameter). The two different external magnetic fields were applied by combining a permanent magnet and an electric magnetic coil connected in an H-bridge configuration, which can reverse the current direction in the coil. A BioScope Resolve AFM (Bruker) was used for cell surface imaging and for nanoindentation. The spatial correlation between the AFM and the confocal microscope was established by overlapping the AFM image and the fluorescence image of NDs on the substrate. The AFM provides the coordinates of different indentation spots in the scanning probe measurements.

**Cell culture and treatments**

HeLa cells (ATCC) were cultured in a confocal dish with microwave antenna at 37 °C in a 5% $CO_2$ atmosphere. The growth medium was Dulbecco's modified Eagle's medium (DMEM, Gibco) supplemented with 10% fetal bovine serum (Gibco) and 1% penicillin–streptomycin (Gibco).

**Sample preparation**

Cells were cultured for 48 h in an incubator. Before experiments, we washed the sample with PBS for three times and then added 4% paraformaldehyde (PFA) solution in PBS to fix the cells. The fixation process was held for 15 min. at 4 ℃ and finally the PFA solution was removed by PBS rinsing three times. 20 µL ND aqueous solution ( 2µg mL$^{-1}$ concentration with typical 500 NV centers per ND, from Adámas Nanotechnologies) were then drop-casted onto the sample and water was subsequently added to the confocal dish. The dish was mounted on the stage of AFM for the following AFM nanoindentation and ODMR measurements of NDs. For experiments on CytD-treated cells, cells were treated with 5µg mL$^{-1}$ Cytochalasin D solution in DMEM for 40 min. before the PFA fixation.

For visualization of F-actin in cells, after fixation and permeabilization (0.1% Triton™ X-100 in PBS, 15 min.), cells were stained for 50 min. with phalloidin. The fluorescence image of the sample was then recorded on a confocal microscope excited by a 473 nm laser.

**AFM nanoindentation**

All AFM measurements were performed using a BioScope Resolve AFM (Bruker). A conical PFQNM-LC-A probe (calibrated spring constant: 0.089 N m$^{-1}$, for scanning electron microscopy (SEM) image see Supplementary Fig. 1) was used to image the cell surface in a peak force tapping mode and to apply nanoindentation. The indentations were implemented in a so-called constant force mode (ramp and hold trigger force), where a loading force was applied (with indentation rate 600 nm s$^{-1}$, see Supplementary Fig. 3b) and kept constant after reaching the set trigger force (around 10 nN). The cell deformation then increases with time until reaching stabilization (in 10-20 seconds) due to the viscoelastic creep of cell (see Supplementary Fig. 3c). The ODMR spectra of the NDs were collected when the deformation was approximately stable (typically 10-20 seconds after the force being established). To avoid possible artefacts induced by the interaction between NDs and the AFM tip, the indentation positions were chosen not too close to the NDs (typically > 400 nm).

**ND rotation sensing and deformation reconstruction**

The normalized ODMR spectra under the two known magnetic fields are fitted by

$$S(f) = b - \sum_n C_n \left[ \frac{\Delta f^2}{4(f - f_n^+)^2 + \Delta f^2} + \frac{\Delta f^2}{4(f - f_n^-)^2 + \Delta f^2} \right] \quad (1)$$

where $b$ is the baseline, $C_n$ represents the ODMR contrast of the NV centers along the $n$th crystallographic orientation, $\Delta f$ is the linewidth (FWHM), and $f_n^\pm$ are the transition frequencies. We fitted the ODMR spectra under the two magnetic fields before and after the rotation, with the fitting parameters being the Euler angles of the initial ND orientation, the baselines, the contrasts, the FWHM, and the zero-field splitting $D$ (slightly different among different NDs). Then, the rotation of the ND was determined by the two sets of the Euler angles. For details of the fitting methods and results, see [20] and Supplementary Fig. 4.

The rotation of the docked NDs on the deformed surface $z(\boldsymbol{\rho})$ upon indentation is geometrically related to the gradient field as $\mathbf{g}(\boldsymbol{\rho}) = \boldsymbol{\nabla}_{\boldsymbol{\rho}} z(\boldsymbol{\rho}) = \left( -\hat{n}_x / \hat{n}_z, -\hat{n}_y / \hat{n}_z \right)$, where $\hat{\mathbf{n}}$ is the normal vector of the deformed surface. The normal vector was measured by the rotation of the NDs as $\hat{\mathbf{n}} = \mathbf{R}(\theta, \phi, \chi)\hat{\mathbf{z}}$, where $\mathbf{R}$ is the rotation matrix with rotation axis $(\theta, \phi)$ and rotation angle $\chi$, and $\hat{\mathbf{z}}$ is the unit vector along the z-axis. The deformed surface was reconstructed by the least-square fitting of the gradient field (Fig. 2e), in which the missing gradient field was estimated by the curvature-minimizing interpolation [43]. By applying the axisymmetric condition as in our experiments, the no-local deformation was reconstructed by integration $z(\rho) = \int_0^\rho \tan \chi(\rho') d\rho'$.

**Non-local deformation with surface tension effect and data analysis**

For deformation far from the contact area, according to Saint-Veniant's principle [29], the contact model can be approximated to be a point loading $P$ (which means loading at a single point with zero contact size) acting on the surface of an elastic solid with surface tension. The surface tension is introduced by adding a thin membrane ideally adhered to the bulk with negligibly thickness [44]. The normal displacement $z$ on the surface is given by Hajji as [30],

$$z(\rho) = -\frac{z_P}{4}\left[H_0\left(\frac{\rho}{s}\right) - Y_0\left(\frac{\rho}{s}\right)\right], \tag{2}$$

where $\rho$ is the distance from the indentation points, and $H_n$ and $Y_n$ are the Struve function and the Bessel function of the second kind, respectively. The non-local deformation of the soft solid is then uniquely determined by the deformation amplitude $z_P \equiv P/\tau_0$ and the elastocapillary length $s \equiv 2\tau_0/E^*$, where $E^*$ is the elastic modulus including both bulk and shear moduli. The elastic modulus is usually used in contact mechanics and called as effective Young's modulus in literature [12]. The rotation angle of the ND is approximated to be the derivative of the surface profile in the small deformation limit as,

$$\chi(\rho) \approx \partial_\rho z(\rho) = \frac{z_P}{s} f\left(\frac{\rho}{s}\right), \tag{3}$$

where $f(r) = [H_1(r) - H_{-1}(r)]/8 - Y_1(r)/4 - 1/(4\pi)$ (the Hajji function, see Fig. 1c). Under a given loading force $P$, the surface tension $\tau_0$ and the elastic modulus $E^*$ are deduced by the least-square fitting of the rotation angle with Eq. (3).

**The Hertz-Sneddon model**

The indentation with a finite-size parabola tip into a homogenous elastic material is simulated using the Hertz-Sneddon model [28]. The non-local surface profile (Fig. 2i) is written as

$$z(\rho) = \frac{d}{\pi}\left[\left(2 - \frac{\rho^2}{a^2}\right)\sin^{-1}\left(\frac{a}{\rho}\right) + \frac{\rho}{a}\sqrt{1 - \frac{a^2}{\rho^2}}\right], \quad \rho \geq a, \tag{4}$$

where $d$ is the indentation depth and $a = \sqrt{Rd}$ is the contact radius with $R$ being the radius of the parabola tip ($R = 35$ nm as measured by SEM imaging in Supplementary Fig. 1). The rotation angle of the docked ND (Fig. 2h) was calculated using the geometric relation $\chi = \partial_\rho z$. Moreover, the local depth-loading profile is given by,

$$P = \frac{4}{3} E_a^* R^{1/2} d^{3/2}, \tag{5}$$

where $E_a^*$ is the apparent elastic modulus. Therefore, $E_a^*$ (Fig. 3c) was deduced by fitting the local depth-loading curve using Eq. (5).

**Data availability.** The data which supports the findings of this work is available upon request from the corresponding authors.

# Reference

[1]  M. P. Stewart, J. Helenius, Y. Toyoda, S. P. Ramanathan, D. J. Muller, and A. A. Hyman, *Hydrostatic Pressure and the Actomyosin Cortex Drive Mitotic Cell Rounding*, Nature **469**, 226 (2011).

[2]  A. Diz-Muñoz, D. A. Fletcher, and O. D. Weiner, *Use the Force: Membrane Tension as an Organizer of Cell Shape and Motility*, Trends in Cell Biology **23**, 47 (2013).

[3]  P. Sens and J. Plastino, *Membrane Tension and Cytoskeleton Organization in Cell Motility*, Journal of Physics: Condensed Matter **27**, 273103 (2015).

[4]  O. Chaudhuri and D. J. Mooney, *Anchoring Cell-Fate Cues*, Nature Materials **11**, 7 (2012).

[5]  E. E. Charrier, K. Pogoda, R. G. Wells, and P. A. Janmey, *Control of Cell Morphology and Differentiation by Substrates with Independently Tunable Elasticity and Viscous Dissipation*, Nature Communications **9**, 1 (2018).

[6]  M. Lekka, *Discrimination Between Normal and Cancerous Cells Using AFM*, BioNanoSci. **6**, 65 (2016).

[7]  Q. S. Li, G. Y. H. Lee, C. N. Ong, and C. T. Lim, *AFM Indentation Study of Breast Cancer Cells*, Biochemical and Biophysical Research Communications **374**, 609 (2008).

[8]  C. Alibert, B. Goud, and J.-B. Manneville, *Are Cancer Cells Really Softer than Normal Cells?*, Biology of the Cell **109**, 167 (2017).

[9]  N. Guz, M. Dokukin, V. Kalaparthi, and I. Sokolov, *If Cell Mechanics Can Be Described by Elastic Modulus: Study of Different Models and Probes Used in Indentation Experiments*, Biophysical Journal **107**, 564 (2014).

[10] J. Ren, S. Yu, N. Gao, and Q. Zou, *Indentation Quantification for In-Liquid Nanomechanical Measurement of Soft Material Using an Atomic Force Microscope: Rate-Dependent Elastic Modulus of Live Cells*, Phys. Rev. E **88**, 052711 (2013).

[11] E. K. Dimitriadis, F. Horkay, J. Maresca, B. Kachar, and R. S. Chadwick, *Determination of Elastic Moduli of Thin Layers of Soft Material Using the Atomic Force Microscope*, Biophysical Journal **82**, 2798 (2002).

[12] M. Krieg, G. Fläschner, D. Alsteens, B. M. Gaub, W. H. Roos, G. J. L. Wuite, H. E. Gaub, C. Gerber, Y. F. Dufrêne, and D. J. Müller, *Atomic Force Microscopy-Based Mechanobiology*, Nature Reviews Physics **1**, 41 (2019).

[13] Y. M. Efremov, W.-H. Wang, S. D. Hardy, R. L. Geahlen, and A. Raman, *Measuring Nanoscale Viscoelastic Parameters of Cells Directly from AFM Force-Displacement Curves*, Sci Rep **7**, 1541 (2017).

[14] Yu. M. Efremov, D. V. Bagrov, M. P. Kirpichnikov, and K. V. Shaitan, *Application of the Johnson–Kendall–Roberts Model in AFM-Based Mechanical Measurements on Cells and Gel.*, Colloids and Surfaces B: Biointerfaces **134**, 131 (2015).

[15] I. Sokolov, M. E. Dokukin, and N. V. Guz, *Method for Quantitative Measurements of the Elastic Modulus of Biological Cells in AFM Indentation Experiments*, Methods **60**, 202 (2013).

[16] J. T. Pham, F. Schellenberger, M. Kappl, and H.-J. Butt, *From Elasticity to Capillarity in Soft Materials Indentation*, Physical Review Materials **1**, 015602 (2017).

[17] R. W. Style, A. Jagota, C.-Y. Hui, and E. R. Dufresne, *Elastocapillarity: Surface Tension and the Mechanics of Soft Solids*, Annu. Rev. Condens. Matter Phys. **8**, 99 (2017).

[18] Q. Xu, K. E. Jensen, R. Boltyanskiy, R. Sarfati, R. W. Style, and E. R. Dufresne, *Direct Measurement of Strain-Dependent Solid Surface Stress*, Nature Communications **8**, 555 (2017).

[19] K. E. Jensen, R. W. Style, Q. Xu, and E. R. Dufresne, *Strain-Dependent Solid Surface Stress and the Stiffness of Soft Contacts*, Phys. Rev. X **7**, 041031 (2017).

[20] K. Xia, C.-F. Liu, W. Leong, M. Kwok, Z. Yang, X. Feng, R.-B. Liu, and Q. Li, *Nanometer-Precision Non-Local Deformation Reconstruction Using Nanodiamond Orientation Sensing*, Nature Communications **10**, 3259 (2019).

[21] F. Neugart, A. Zappe, F. Jelezko, C. Tietz, J. P. Boudou, A. Krueger, and J. Wrachtrup, *Dynamics of Diamond Nanoparticles in Solution and Cells*, Nano Lett. **7**, 3588 (2007).

[22] L. P. McGuinness, Y. Yan, A. Stacey, D. A. Simpson, L. T. Hall, D. Maclaurin, S. Prawer, P. Mulvaney, J. Wrachtrup, F. Caruso, R. E. Scholten, and L. C. L. Hollenberg, *Quantum Measurement and Orientation Tracking of Fluorescent Nanodiamonds inside Living Cells*, Nature Nanotechnology **6**, 6 (2011).

[23] G. Kucsko, P. C. Maurer, N. Y. Yao, M. Kubo, H. J. Noh, P. K. Lo, H. Park, and M. D. Lukin, *Nanometre-Scale Thermometry in a Living Cell*, Nature **500**, 54 (2013).

[24] D. Le Sage, K. Arai, D. R. Glenn, S. J. DeVience, L. M. Pham, L. Rahn-Lee, M. D. Lukin, A. Yacoby, A. Komeili, and R. L. Walsworth, *Optical Magnetic Imaging of Living Cells*, Nature **496**, 7446 (2013).

[25] R. Igarashi, T. Sugi, S. Sotoma, T. Genjo, Y. Kumiya, E. Walinda, H. Ueno, K. Ikeda, H. Sumiya, H. Tochio, Y. Yoshinari, Y. Harada, and M. Shirakawa, *Tracking the 3D Rotational Dynamics in Nanoscopic Biological Systems*, Journal of the American Chemical Society **142**, 7542 (2020).

[26] M. S. J. Barson, P. Peddibhotla, P. Ovartchaiyapong, K. Ganesan, R. L. Taylor, M. Gebert, Z. Mielens, B. Koslowski, D. A. Simpson, L. P. McGuinness, J. McCallum, S. Prawer, S. Onoda, T. Ohshima, A. C. Bleszynski Jayich, F. Jelezko, N. B. Manson, and M. W. Doherty, *Nanomechanical Sensing Using Spins in Diamond*, Nano Letters **17**, 1496 (2017).

[27] M. Targosz-Korecka, G. D. Brzezinka, J. Danilkiewicz, Z. Rajfur, and M. Szymonski, *Glutaraldehyde Fixation Preserves the Trend of Elasticity Alterations for Endothelial Cells Exposed to TNF-α*, Cytoskeleton **72**, 124 (2015).

[28] I. N. Sneddon, *The Relation between Load and Penetration in the Axisymmetric*

*Boussinesq Problem for a Punch of Arbitrary Profile*, International Journal of Engineering Science **3**, 47 (1965).

[29] A. E. H. Love, *A Treatise on the Mathematical Theory of Elasticity* (New York : Dover Publications, 1944, n.d.).

[30] M. A. Hajji, *Indentation of a Membrane on an Elastic Half Space*, Journal of Applied Mechanics **45**, 320 (1978).

[31] Y. Ding, J. Wang, G.-K. Xu, and G.-F. Wang, *Are Elastic Moduli of Biological Cells Depth Dependent or Not? Another Explanation Using a Contact Mechanics Model with Surface Tension*, Soft Matter **14**, 7534 (2018).

[32] M. Lekka, D. Gil, K. Pogoda, J. Dulińska-Litewka, R. Jach, J. Gostek, O. Klymenko, S. Prauzner-Bechcicki, Z. Stachura, J. Wiltowska-Zuber, K. Okoń, and P. Laidler, *Cancer Cell Detection in Tissue Sections Using AFM*, Archives of Biochemistry and Biophysics **518**, 151 (2012).

[33] E. Fischer-Friedrich, A. A. Hyman, F. Jülicher, D. J. Müller, and J. Helenius, *Quantification of Surface Tension and Internal Pressure Generated by Single Mitotic Cells*, Scientific Reports **4**, 6213 (2015).

[34] P. Chugh, A. G. Clark, M. B. Smith, D. A. D. Cassani, K. Dierkes, A. Ragab, P. P. Roux, G. Charras, G. Salbreux, and E. K. Paluch, *Actin Cortex Architecture Regulates Cell Surface Tension*, Nature Cell Biology **19**, 689 (2017).

[35] H. Hubrich, I. P. Mey, B. R. Brückner, P. Mühlenbrock, S. Nehls, L. Grabenhorst, T. Oswald, C. Steinem, and A. Janshoff, *Viscoelasticity of Native and Artificial Actin Cortices Assessed by Nanoindentation Experiments*, Nano Letters **20**, 6329 (2020).

[36] S.-O. Kim, J. Kim, T. Okajima, and N.-J. Cho, *Mechanical Properties of Paraformaldehyde-Treated Individual Cells Investigated by Atomic Force Microscopy and Scanning Ion Conductance Microscopy*, Nano Convergence **4**, (2017).

[37] H. Liang, Z. Cao, Z. Wang, and A. V. Dobrynin, *Surface Stress and Surface Tension in Polymeric Networks*, ACS Macro Letters **7**, 116 (2018).

[38] S. Leporatti, D. Vergara, A. Zacheo, V. Vergaro, G. Maruccio, R. Cingolani, and R. Rinaldi, *Cytomechanical and Topological Investigation of MCF-7 Cells by Scanning Force Microscopy*, Nanotechnology **20**, 055103 (2009).

[39] A. Berquand, C. Roduit, S. Kasas, A. Holloschi, L. Ponce, and M. Hafner, *Atomic Force Microscopy Imaging of Living Cells*, Micros. Today **18**, 8 (2010).

[40] M. Schliwa, *Action of Cytochalasin D on Cytoskeletal Networks*, The Journal of Cell Biology **92**, 79 (1982).

[41] C. Rotsch and M. Radmacher, *Drug-Induced Changes of Cytoskeletal Structure and Mechanics in Fibroblasts: An Atomic Force Microscopy Study*, Biophysical Journal **78**, 520 (2000).

[42] A. R. Harris and G. T. Charras, *Experimental Validation of Atomic Force Microscopy-Based Cell Elasticity Measurements*, Nanotechnology **22**, 345102 (2011).

[43] G. M. Nielson, *A Method for Interpolating Scattered Data Based Upon a Minimum Norm Network*, Mathematics of Computation **40**, 253 (1983).

[44] M. E. Gurtin and A. Ian Murdoch, *A Continuum Theory of Elastic Material Surfaces*, Archive for Rational Mechanics and Analysis **57**, 291 (1975).

## Acknowledgements

R.B.L. acknowledges the funding support by RGC/CRF of HKSAR under project no. C4007-19G. Q.L. and C.G. acknowledge the funding support from ANR/RGC Joint Research Project under project no. AA-PG-176343/ A-CUHK404/18.

## Author contributions

Q. L. and R. B. L. conceived the idea and supervised the project. W. H. L, Y. C., R. B. L. and Q. L. designed the experiments. Y. C. performed the experiments. W. H. L., Y. C., R. B. L. and Q. L. analyzed the data. C.G., Q. L., and Y. C. discussed preparation of cell samples for indentation and sensing experiments. Y. C. and X. F. prepared the samples. C. F. L. and K. X. carried out the preliminary experiments. Y. C., W. H. L., R. B. L. and Q. L. wrote the paper and all authors commented on the manuscript.

**Competing interests.** The authors declare no competing interests.

**Materials & Correspondence.** Correspondence and requests for materials should be addressed to R.-B.L. (email: rbliu@cuhk.edu.hk) or to Q.L. (email: liquan@cuhk.edu.hk)

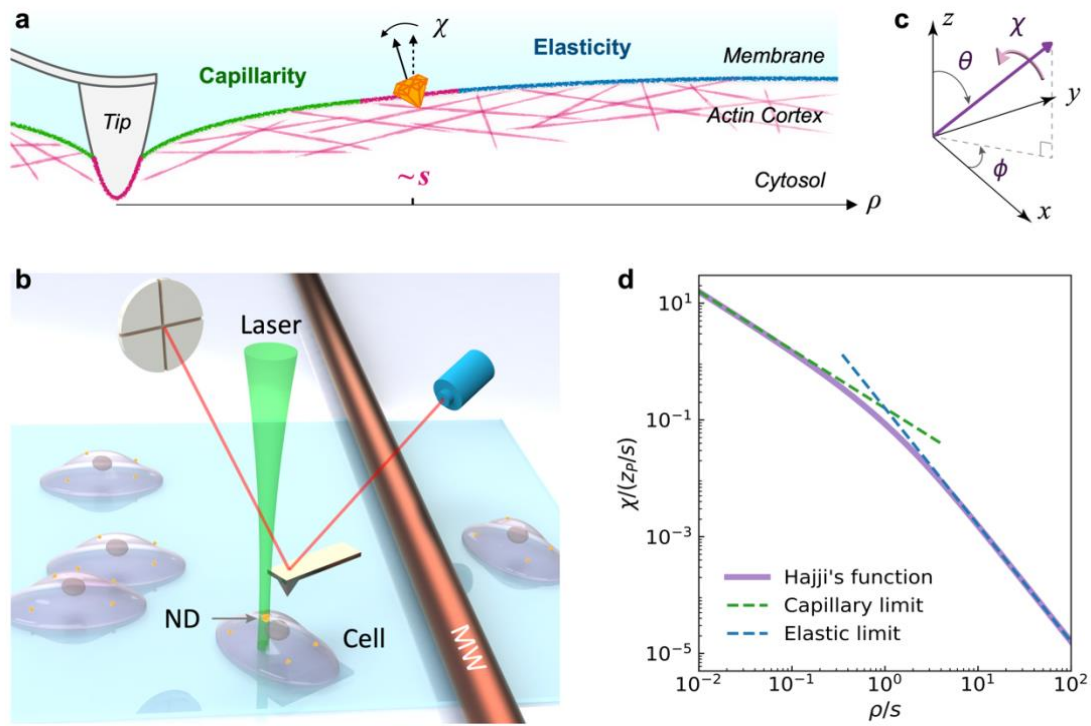

**Figure 1 | Simultaneous evaluation of the capillarity and elasticity of cell surface/subsurface by nanodiamonds (NDs) rotation sensing . a** Schematic of an atomic force microscopy (AFM) indentation on cell surface/subsurface structure. The non-local deformation at distance $\rho$ is governed by the interplay between capillary forces and elasticity near the elastocapillary length $s$. The ND docked on the surface rotates due to the indentation induced deformation of the cell. **b** Schematic of an AFM tip imposing a localized indentation on a fixed cell. NDs are attached on the surface of the cell with a nearby microwave antenna for rotation sensing measurement. **c** Rotation of the docked ND is characterized by the direction of the rotation axis $(\theta, \phi)$ and the rotation angle $\chi$. **d** The normalized rotation angle of the docked ND upon a point loading as a function of $\rho/s$ (the Hajji function, see the text). The dashed green (blue) line shows the asymptotic from in the capillary (elastic) limit while $\rho \ll s$ ($\rho \gg s$).

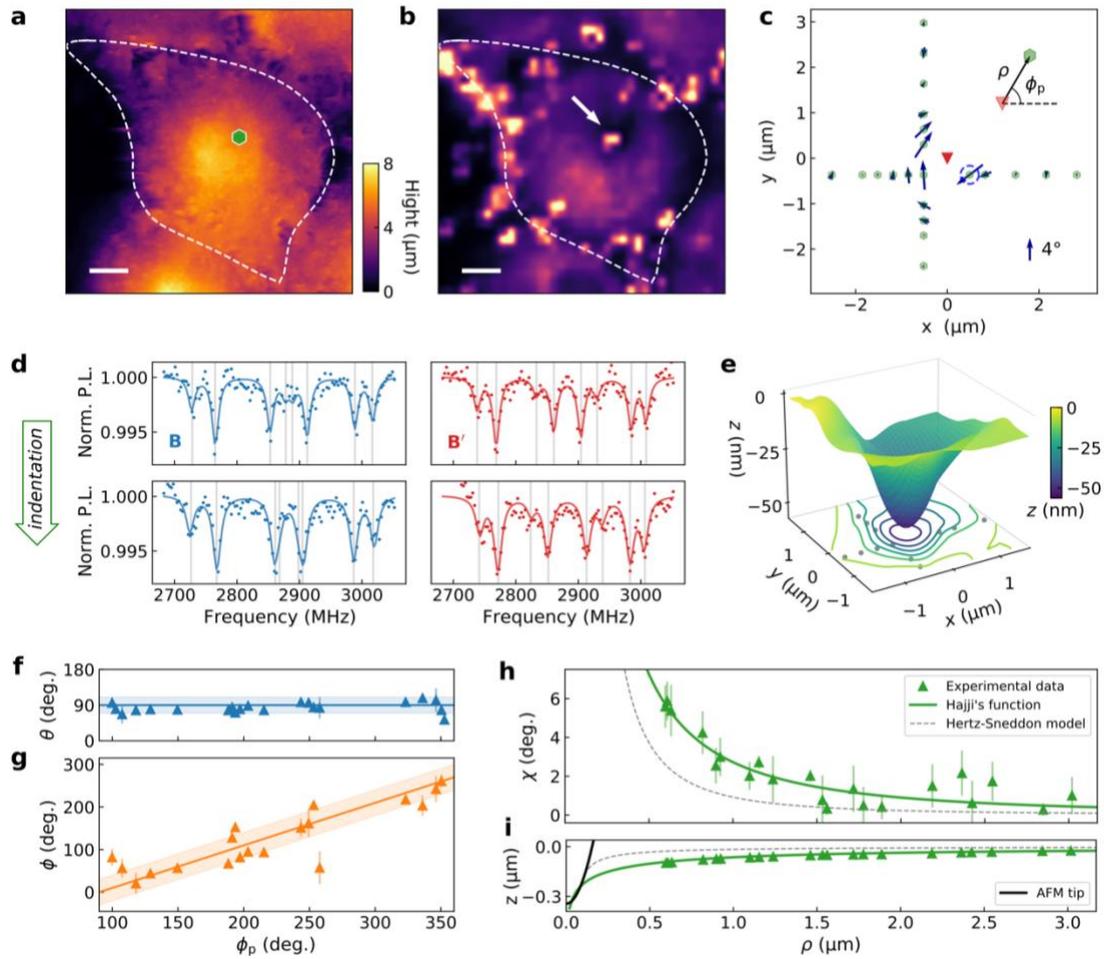

**Figure 2 | Deformation reconstruction using rotation data of a single-nanodiamond on a fixed HeLa cell. a** The AFM image and **b** the fluorescence confocal image of a fixed cell (enclosed by the white dashed lines). The green hexagon in **a** and the white arrow in **b** indicate the position of the ND. The scale bars in **a** and **b** are 5 $\mu m$. **c** The rotation of the ND (represented by arrows, with arrow direction representing the projected rotation axis on the x-y plane, and the arrow length proportional to the magnitude of rotation angle) plotted as a function of displacement $\rho$ of the ND from the indentation positions (red triangle). **d** ODMR spectra (dots) obtained for the indentation location marked in **c** (dashed blue circle) under the external magnetic fields $B$ (blue) and $B'$ (red) with and without the indentation (upper vs. lower panel). The

solid lines are the fitting results. The vertical grey lines mark the resonance frequencies. **e** The reconstructed surface of the cell upon indentation. **f** The polar angle $\theta$ and **g** the azimuthal angle $\phi$ defining the ND rotation axes as functions of $\phi_p$. The solid line in **f** is $\theta = 90°$ and in **g** is $\phi = \phi_p - 90°$, with the shadowed regions bounded by $\pm 20°$ and $\pm 30°$. **h** The rotation angle $\chi$ and **i** the reconstructed deformation as functions of the distance $\rho$. The green solid lines show the best fitted Hajji function and the corresponding surface profile. The dashed grey lines are the simulation results using the Hertz-Sneddon model with elasticity only. The vertical error bars in **f**, **g**, and **h** are fitting errors.

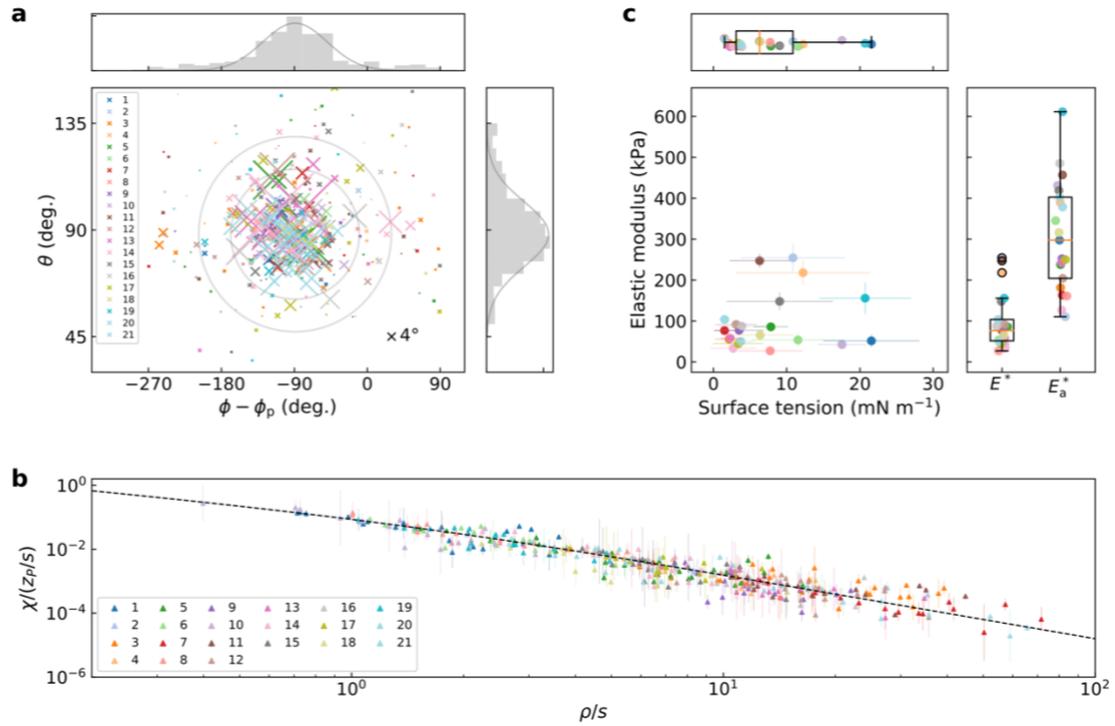

**Figure 3 | Mechanical property evaluation of fixed HeLa cells. a** The distribution of the rotation axes $(\theta, \phi)$ of docked nanodiamonds (NDs) on the surfaces of twenty one fixed HeLa cells measured (represented by different colors 1-21) under atomic force microscopy (AFM) indentations (The schematic of indentations on each cell can be found in Supplementary Figs. 3a and 5). The azimuthal angle of the rotation axis is plotted as $\Delta\phi = \phi - \phi_p$, where $\phi_p$ is the angle of $\boldsymbol{\rho}$ in the polar coordinates. The size of the color crosses represents the magnitude of the rotation angle $\chi$ (scale bar shown on the lower right). The upper and right graphs are the histograms of the angle $\theta$ and $\Delta\phi$, respectively. The grey solid lines are the Gaussian fitting. The grey circles in the main panel are the 1-, 2- and 3-sigma ellipses of the multivariate Gaussian distribution. **b** Normalized rotation angle $\chi$ as functions of the rescaled distance $\rho/s$. The black dashed lines are the Hajji function (see the text). **c** The surface tension $\tau_0$

and the elastic modulus $E^*$ of the cell surfaces. The upper and right graphs are the box plots of $\tau_0$ and $E^*$. The apparent elastic moduli $E_a^*$ evaluated from the local deformation using the Hertz-Sneddon model are shown as the box plot in the right graph. The error bars in **b** and **c** are fitting errors.

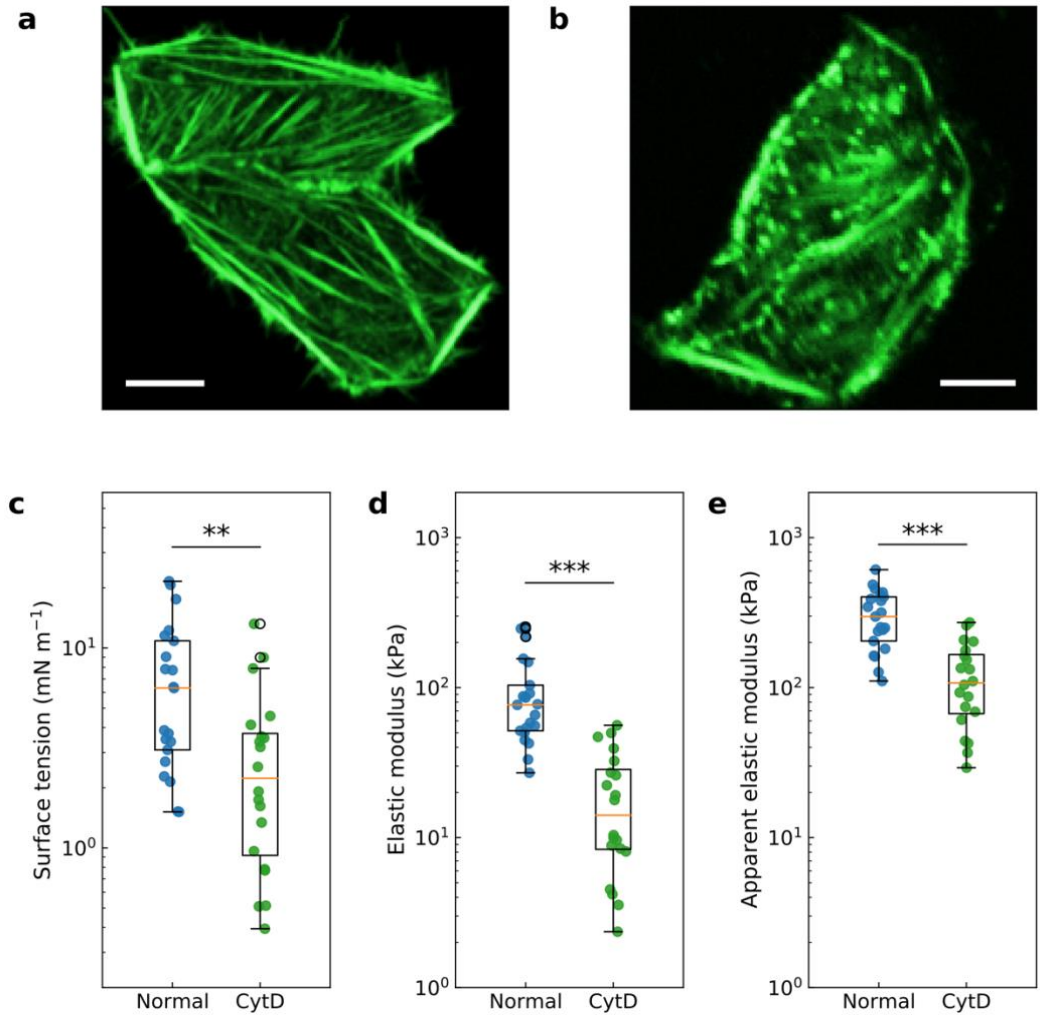

**Figure 4 | Effect of Cytochalasin D (CytD) treatment on the mechanical properties of HeLa cells. a** and **b** Fluorescence confocal images of a HeLa cell and a cell treated with CytD, both stained with phalloidin. The scale bars in **a** and **b** are $10\ \mu m$. Box plots comparing **c** surface tension, **d** elastic modulus, and **e** apparent elastic modulus between HeLa cells and cells treated with CytD (n=20 and 21 cells; $P = 0.008$, $5.7 \times 10^{-5}$ and $4 \times 10^{-6}$, respectively). Welch's *t*-test *P*-values: **$P < 0.01$; ***$P < 0.001$.

# Supplementary Information

*for*

# Measurement of single-cell elasticity by nanodiamond-sensing of non-local deformation

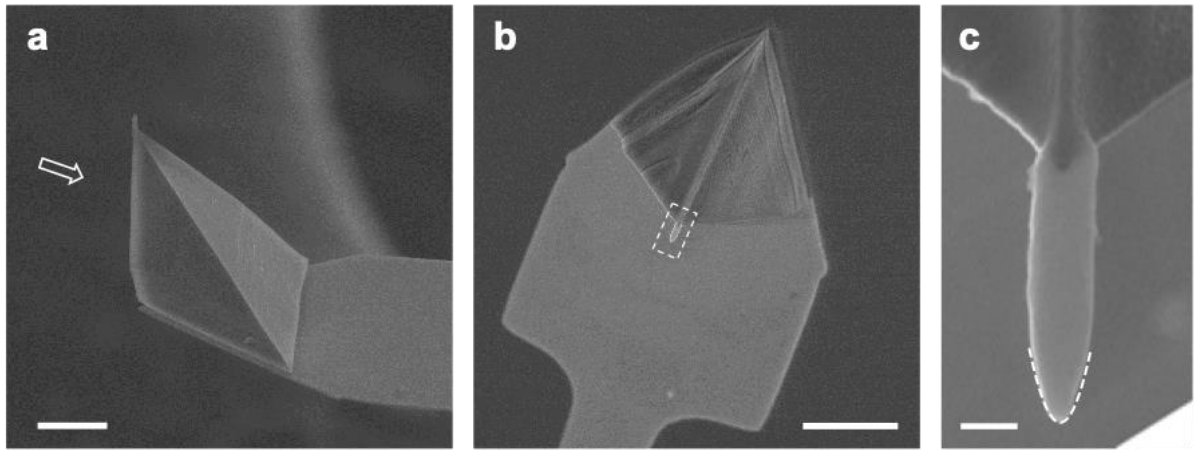

**Supplementary Figure 1 | Scanning electron microscopy (SEM) image of an atomic force microscopy (AFM) tip. a** SEM image of a typical tip (PFQNM-LC-A, Bruker, spring constant: 0.089 N m$^{-1}$) used in the indentation experiments. **b** The SEM image by looking at the tip in the direction indicated by the white arrow in **a**. The scale bars in **a** and **b** are 5 μm. **c** The zoom-in SEM image of the tip in the region enclosed by the dashed rectangle in **b**. The white dashed line outlines the parabola shape ($y = x^2/(2R)$, where $R = 35$ nm) of the tip used in the simulation. The scale bar is 300 nm.

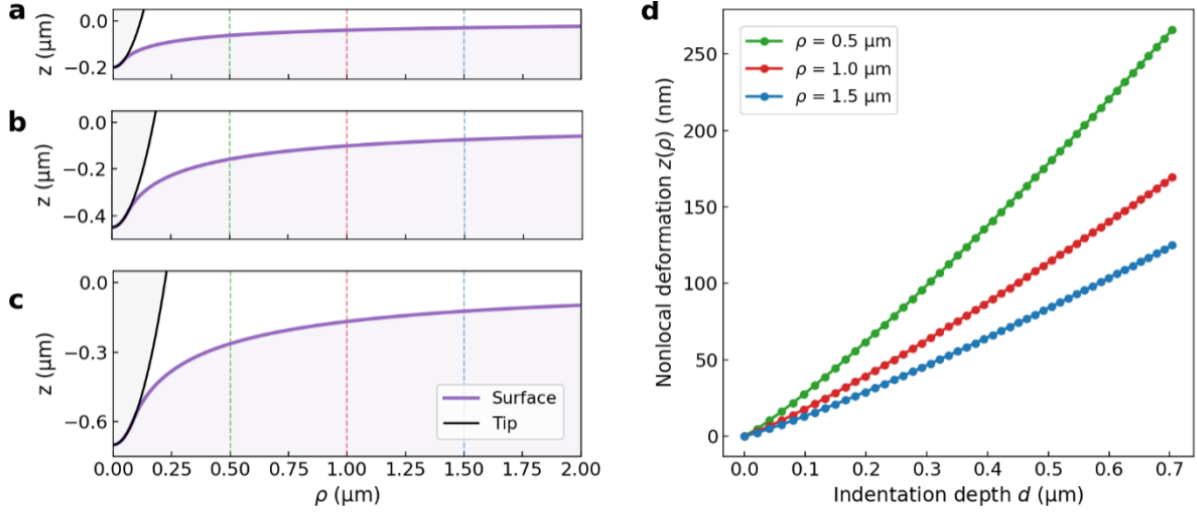

**Supplementary Figure 2 | Simulation of the non-local deformation upon parabola tip indentation with both elasticity and surface tension.** A linear elastic model with finite size tip indentation including the surface tension effect [1] was employed to simulate the deformation of a soft material, which has elastic modulus $E^* = 16$ kPa and surface tension $\tau_0 = 8$ mN m$^{-1}$ (elastocapillary length $s = 1$ μm, similar to those for cells). In the numerical simulation, the equation of equilibrium is solved self-consistently under the given boundary condition (tip shape, surface tension, etc.). The tip is chosen to be parabola and sufficiently small (with radius $R = 35$ nm, consisted with the tip used in the experiment, see Supplementary Fig. 1) to ensure that the contact radius is much more small than $s$. **a** to **c** the simulated surface deformation (the purple lines) with different indentation depth $d = 200, 450$ and 700 nm under the corresponding loading force $P = 2.6, 6.7$ and 11.2 nN, correspondingly. The black lines represent the tips. **d** The nonlocal deformation at different distances ($\rho = 0.5, 1.0$ and 1.5 μm, indicated by the colour dashed lines in **a** to **c**) as function of indentation depth. The deformation is in the range of 40 to 170 nm at $\rho = 1$ μm for indentation depth in the range of 200 to 700 nm.

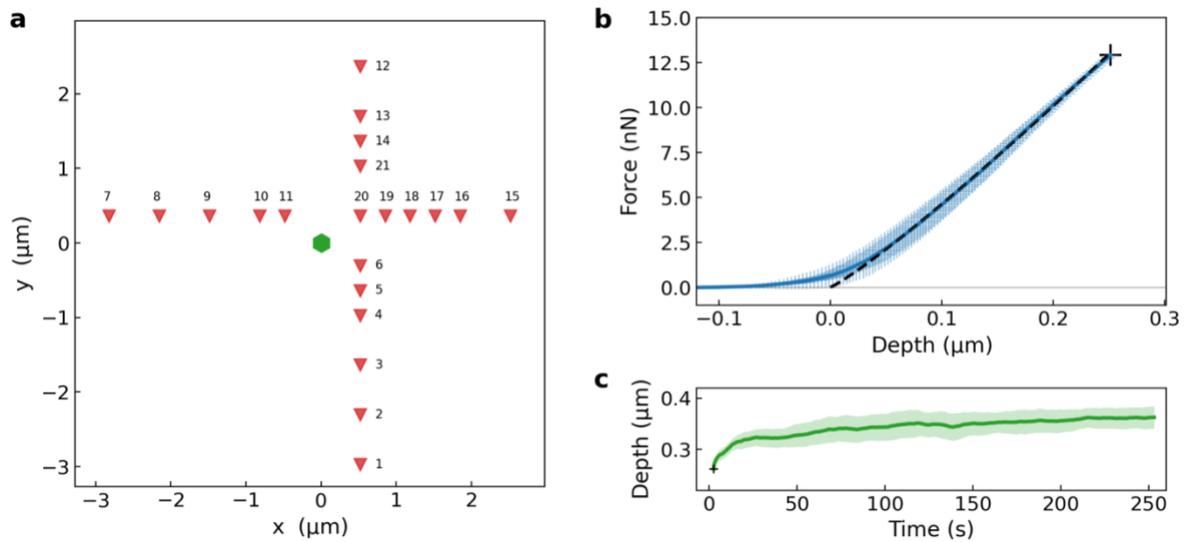

**Supplementary Figure 3 | Local deformation of indentations on a fixed HeLa cell (No.1).**
**a** 21 positions (red triangles) of the AFM tip indentations near a nanodiamond (ND, green hexagon at the origin) on the cell. **b** Averaged depth-loading curve of the 21 local deformations under a constant indentation rate ($600 \text{ nm s}^{-1}$). The vertical drift of the $z$ piezo scanner between different indentations was corrected by aligning and fixing the points (black cross) reaching the specific force set point (e.g., 13 nN for cell No.1). The error bars show the standard deviations of the measurements. To eliminate the effect of the brush layer on the cell surface [2], the zero of the depth was determined by the power-law fitting of the data about 0.2 µm from the fixed points. The power exponent is restricted between 1 and 2 according to the Hertz-Sneddon model [3]. The fitting results are shown by black dashed lines. **c** The averaged depth of the indentations measured by AFM as a function of the holding time with the constant force (green line). The error bars show the standard deviations.

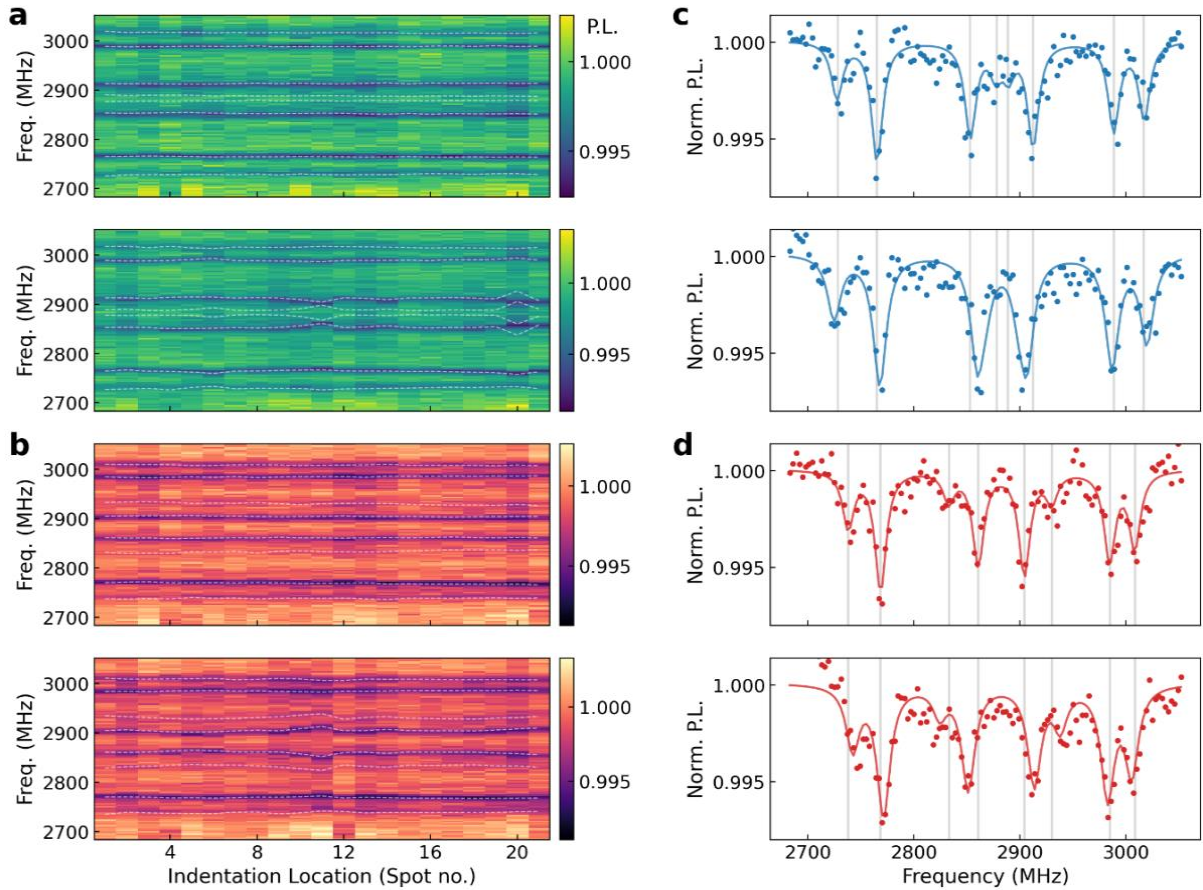

**Supplementary Figure 4 | Optically detected magnetic resonance (ODMR) spectra of the ND on a fixed HeLa cell (No. 1).** ODMR spectra under the external magnetic fields **a B** and **b B′** with (lower panel) and without (upper panel) AFM indentations at different indentation locations (red triangles in Supplementary Fig. 3a). The dashed lines are the fitted resonance frequencies from individual spectra. An example of one set of the original ODMR spectra (dots) taken by indenting at spot 11 are respectively shown in **c** and **d** under the external magnetic fields **B** and **B′** with (lower panel) and without (upper panel) the indentation. The fitting results are shown by solid lines in the same figure. The resonance frequencies of the ODMR spectra without the indentation are marked by grey lines. Significant shifts of the resonance frequencies are observed due to the rotation of the ND upon the indentation. The two magnetic fields **B** and **B′** were calibrated with a bulk diamond before the cell experiments with magnitudes $|\mathbf{B}| = 57.0$ Gauss and $|\mathbf{B'}| = 55.5$ Gauss, and directions $\hat{\mathbf{B}} = (105°, 35°)$ and $\hat{\mathbf{B}}' = (83°, 35°)$. The magnetic fields were calibrated in each set of indentation experiments.

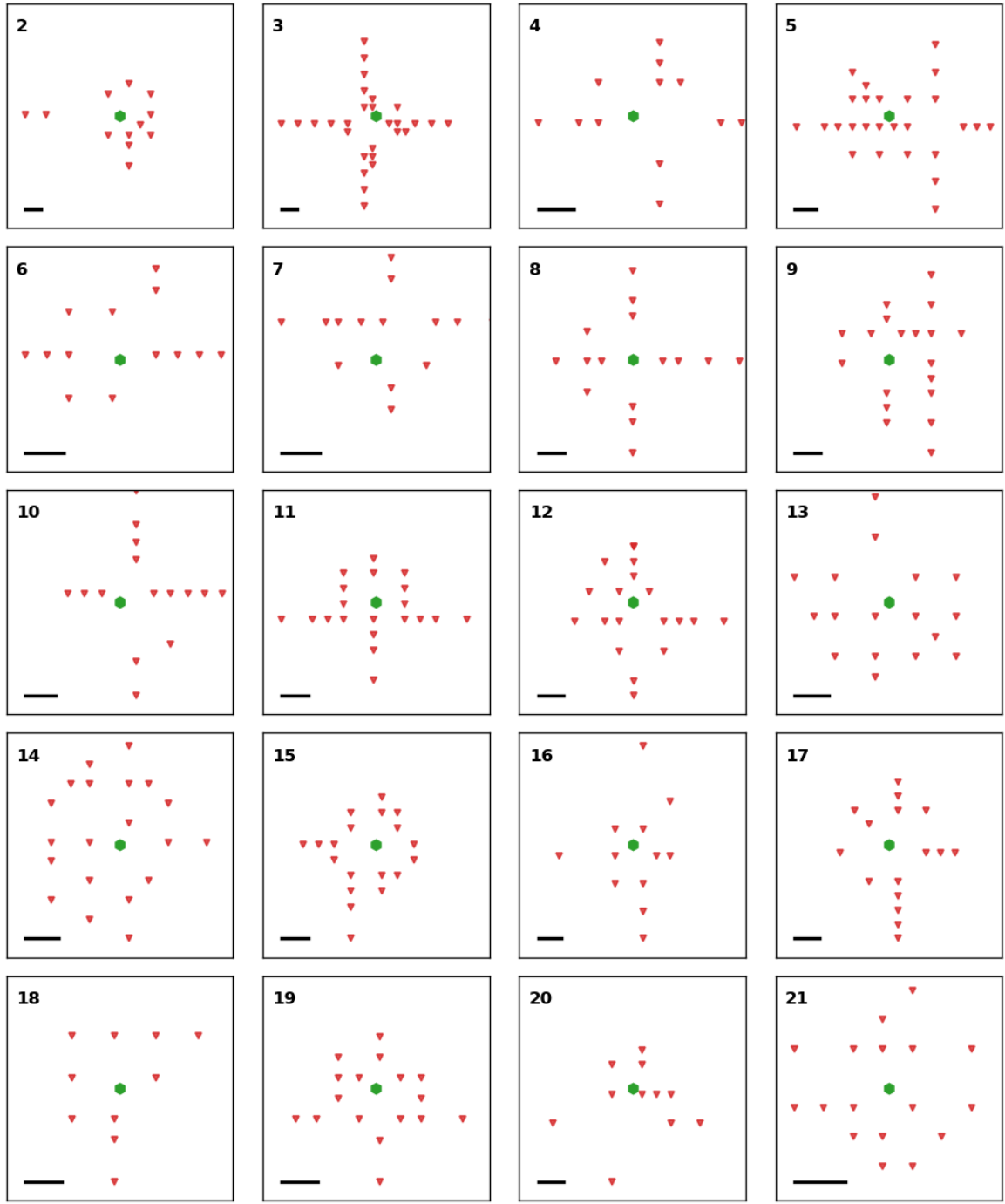

**Supplementary Figure 5 | Indentation positions on twenty HeLa cells (No. 2 to 21).** Positions (red triangles) of the AFM tip indentations near a nanodiamond (ND, green hexagon at the origin) on the cells. The scale bars are 500 nm.

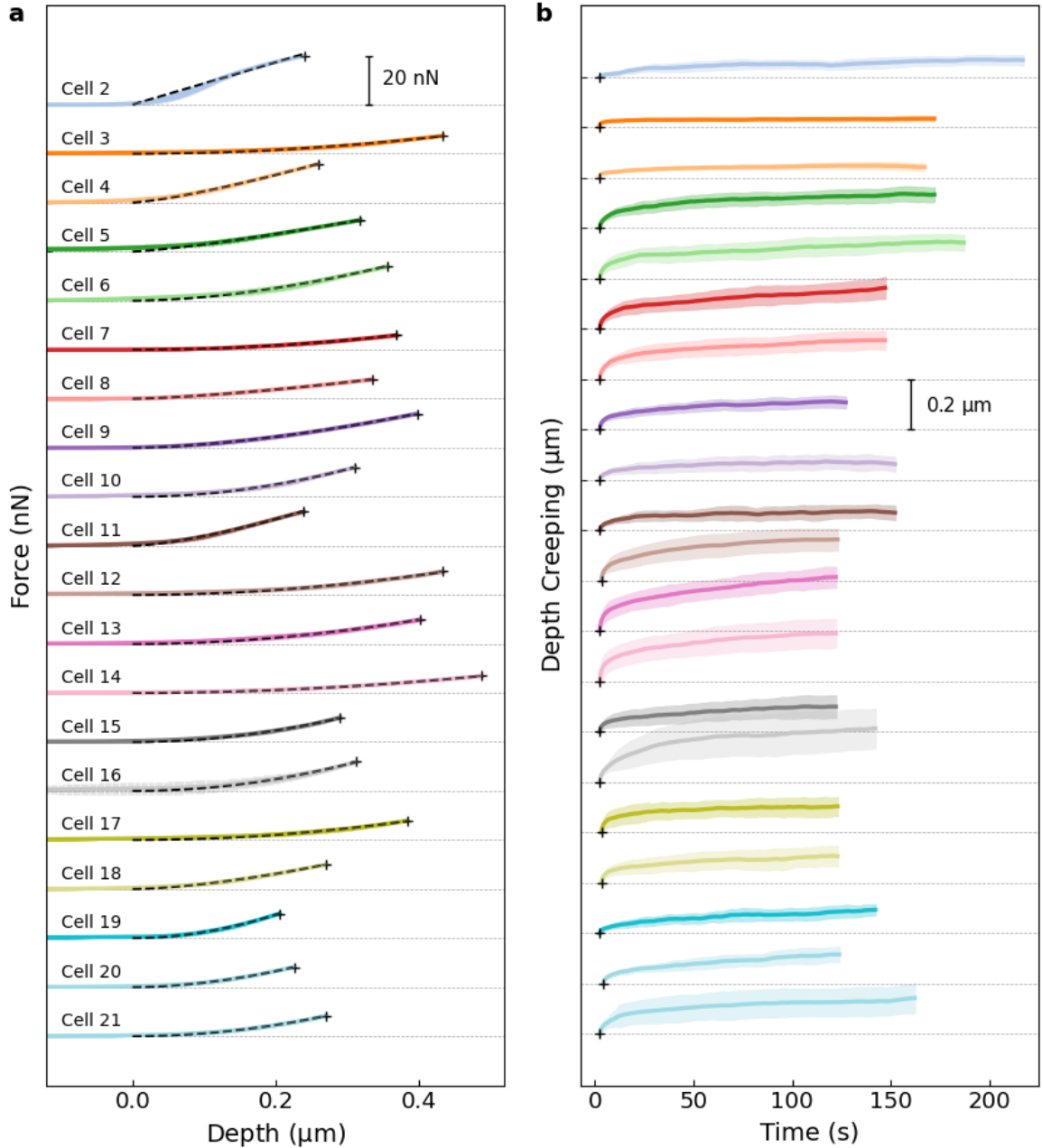

**Supplementary Figure 6 | Local deformation of indentations on HeLa cells No. 2 to 21. a** Averaged depth-loading curves of the deformations on the twenty cells, plot the same way as in Supplementary Fig. 3b. **b** The averaged depths of the indentations as functions of the holding time with the specific constant forces (black crosses in **a**). The error bars show the standard deviations of the measurements.

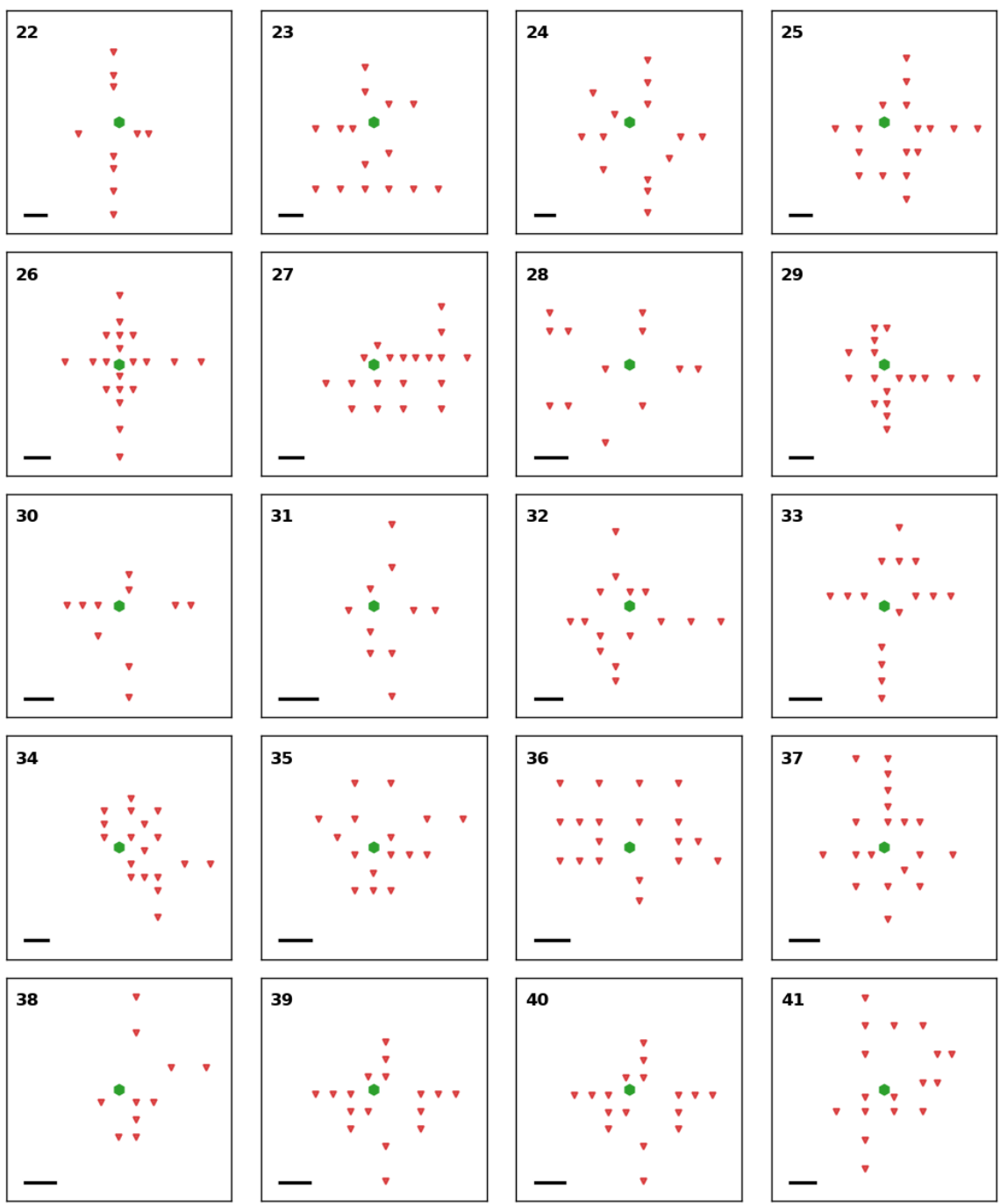

**Supplementary Figure 7 | Indentation positions on twenty HeLa cells treated with CytD (No. 22 to 41).** Positions (red triangles) of the AFM tip indentations near a nanodiamond (ND, green hexagon at the origin) on the cells. The scale bars are 500 nm.

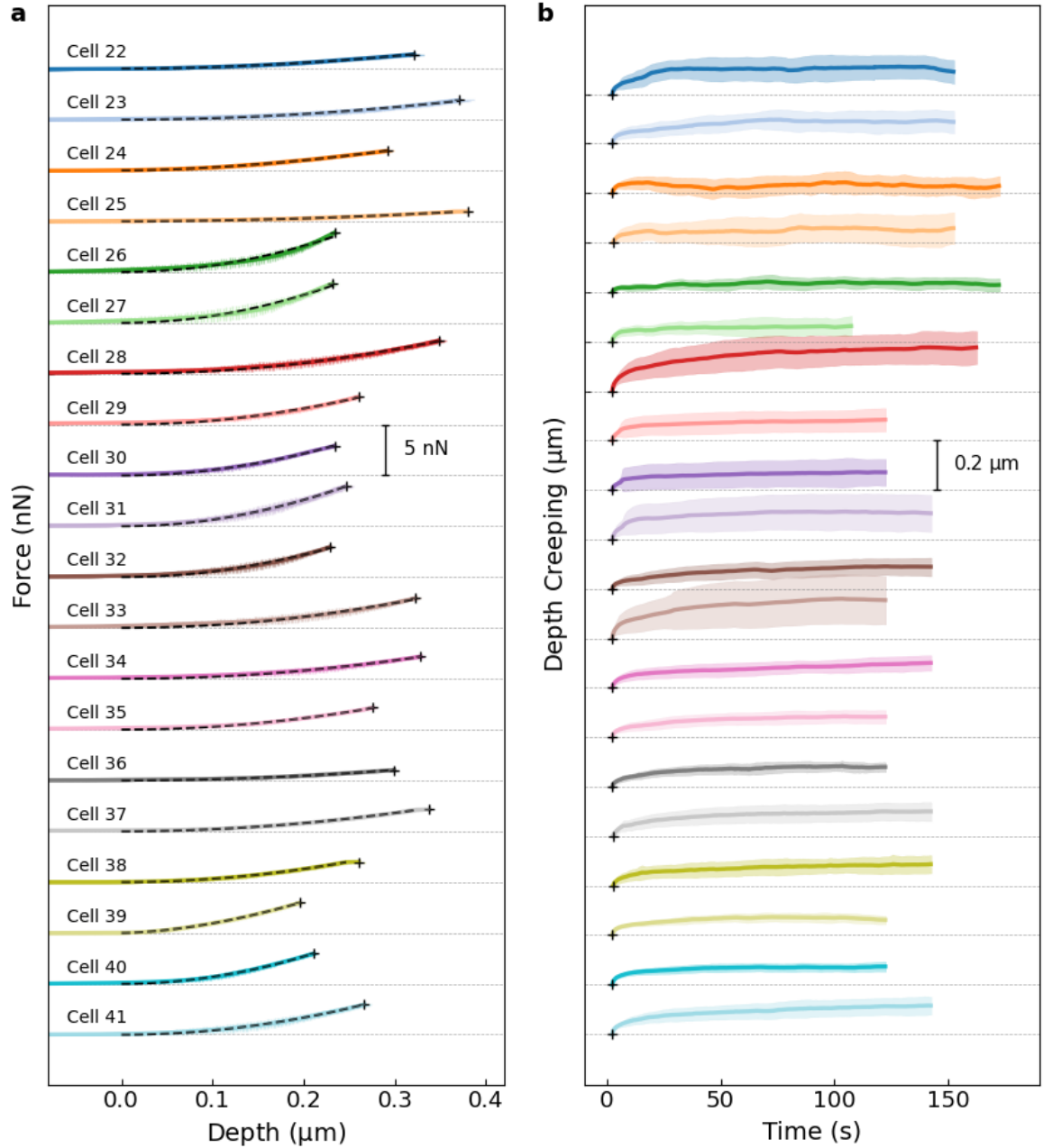

**Supplementary Figure 8 | Local deformations of indentations on the HeLa cells treated with CytD (No. 22-41). a** Averaged depth-loading curve of the deformations on the twenty cells, plot the same way as in Supplementary Fig. 3b. **b** The averaged depths of the indentations as functions of the holding time with the constant force (black crosses in **a**). The error bars show the standard deviations.

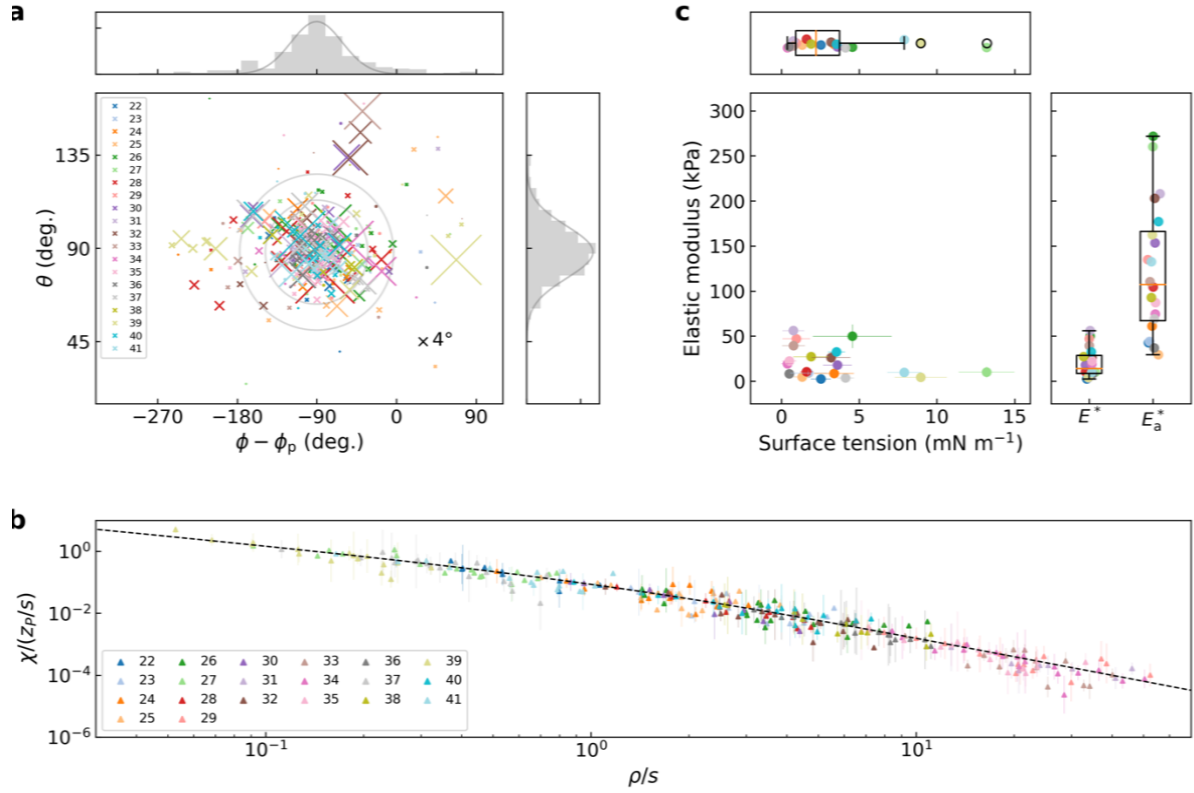

**Supplementary Figure 9 | Mechanical property evaluation of the HeLa cells treated with CytD (No. 22-41). a** The distribution of the rotation axis $(\theta, \phi)$ of the docked NDs taken from the twenty treated cells (represented by coloured crosses for No. 22-41). The azimuthal angle of the rotation axis is plotted as $\Delta\phi = \phi - \phi_p$, where $\phi_p$ is the angle describing the ND location from the AFM tip in a polar coordinate (see Fig. 2c in the main text). The size of the colour crosses represents the magnitude of the rotation angle $\chi$ (scale shown on the lower right). The upper and right graphs are the histograms of the angles and the grey solid lines are the Gaussian function fitting with standard deviation $\sigma_\theta = 13°$ and $\sigma_{\Delta\phi} = 29°$. **b** The normalized rotation angles $\chi$ as functions of the rescaled distance $\rho/s$ between the NDs and the AFM indention positions. The black dashed lines are the best fitted Hajji functions. **c** The surface tension $\tau_0$ and the elastic modulus $E^*$ of the cell surfaces. The upper and right graphs are the box plots of $\tau_0$ and $E^*$. The apparent elastic modulus $E_a^*$ of 20 HeLa cells (No. 22-41) evaluated from the local deformation by using the Hertz-Sneddon model [3] is shown as the box plot in the right graph. The error bars in **b, c** and **d** are the fitting errors.

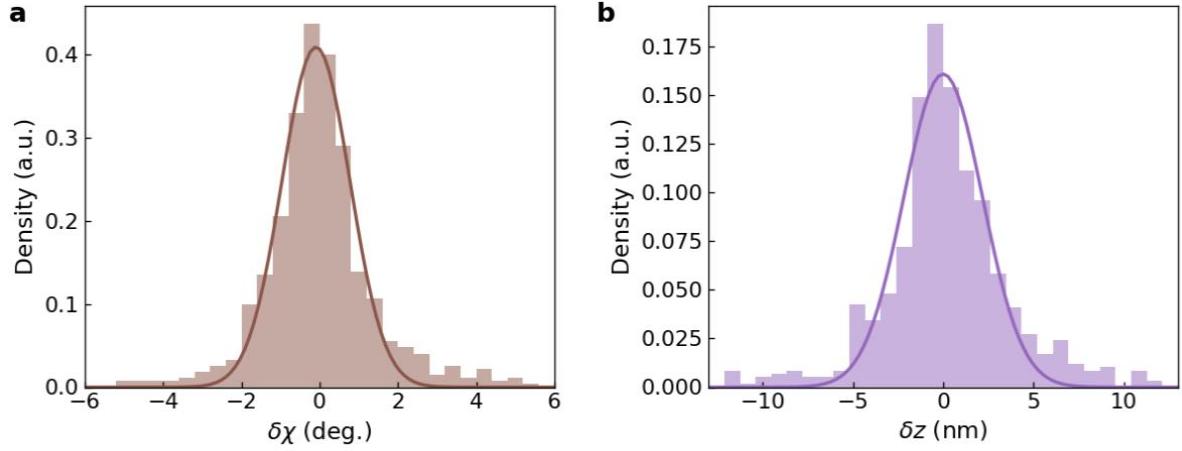

**Supplementary Figure 10 | Comparison between the experimental and simulation results of the HeLa cells (No. 1-41).** The histogram of the deviations between the experimental and theoretical results of **a** the rotation angle $\delta\chi = \chi - \chi_s$ and **b**, the non-local deformation $\delta z = z - z_s$ for the 41 HeLa cells measured in Supplementary Figs. 3, 5 and 7. The experimental results $\chi$ and $z$ were obtained by ODMR measurements and deformation reconstruction, for example see the green dots in Figs. 2h and i in the main text. The theoretical results $\chi_s$ and $z_s$ (such as the green lines in Figs. 2h and i) were calculated by using the Hajji function (see Methods in the main text) and the evaluated mechanical properties (shown in Fig. 3d in the main text). The solid lines are the fitting result with Gaussian distribution, with standard deviations $\sigma_{\delta\chi} = 0.9°$ and $\sigma_{\delta z} = 2$ nm, respectively.

**Reference**

[1] J.M. Long and G.F. Wang, *Effects of Surface Tension on Axisymmetric Hertzian Contact Problem*, Mechanics of Materials **56**, 65 (2013).
[2] Igor Sokolov, Maxim E. Dokukin, and Nataliia V. Guz, *Method for Quantitative Measurements of the Elastic Modulus of Biological Cells in AFM Indentation Experiments*, Methods **60**, 202 (2013).
[3] Ian N. Sneddon, *The Relation between Load and Penetration in the Axisymmetric Boussinesq Problem for a Punch of Arbitrary Profile*, International Journal of Engineering Science **3**, 47 (1965).